\documentclass[12pt]{article}
\usepackage{}
\usepackage{amsfonts,amsmath,amsthm,color,amssymb,graphicx,subfigure,adjustbox}
\usepackage{algorithm,enumerate}

\newcommand{\blind}{1}
\allowdisplaybreaks
\usepackage[authoryear, sort&compress]{natbib} 
\usepackage{bm}
\def\boxit#1{\vbox{\hrule\hbox{\vrule\kern6pt
\vbox{\kern6pt#1\kern6pt}\kern6pt\vrule}\hrule}}

\newcommand{\sbf}{\boldsymbol}

\newtheorem{definition}{Definition}  %
\newtheorem{lemma}{Lemma}            %
\newtheorem{theorem}{Theorem}        %
\newtheorem{proposition}{Proposition}     %
\newtheorem{remark}{Remark}          %
\newtheorem{condition}{Condition}    %

\def\0{{\bf 0}}

\def\be{\begin{eqnarray}}
\def\ee{\end{eqnarray}}

\def\dsum{\sideset{}{_U}{\mathop{\sum\!\sum}}_{k\neq l}}
\def\dsumuu{\sideset{}{_{U_2}}{\mathop{\sum\!\sum}}_{k\neq l}}
\def\dsums{\sideset{}{_s}{\mathop{\sum\!\sum}}_{k\neq l}}
\def\dsumss{\sideset{}{_{s_2}}{\mathop{\sum\!\sum}}_{k\neq l}}
\makeatletter

\newcommand{\Rma}[1]{\textup{\expandafter\@slowromancap\romannumeral #1@}}
\makeatother

\def\tit.arg{\textbf{Improved Horvitz-Thompson Estimator in Survey Sampling}}

\def\abst.arg{
The Horvitz-Thompson (HT) estimator is widely used in survey sampling. However, the variance of the HT estimator becomes large when the inclusion probabilities are highly heterogeneous.
To overcome this shortcoming, in this paper, a hard-threshold method is used for the first-order inclusion probabilities, that is, we carefully choose a threshold value, then replace the inclusion probabilities smaller than the threshold by the threshold. By this shrinkage strategy, we propose a new estimator called improved Horvitz-Thompson (IHT) estimator to estimate the population total.
The IHT estimator increases the estimation accuracy although it brings bias which is relatively small.
We derive the IHT estimator's MSE and its unbiased estimator, and theoretically compare the IHT estimator with the HT estimator. We also apply our idea to construct the improved ratio estimator.
 We numerically analyze simulated and real data sets  to illustrate that the proposed  estimators are more efficient and robust than the classical estimators.

}

\def\key.arg{Horvitz-Thompson estimator; Ratio estimator; Robustness; Unequal probability sampling; Sampling without replacement;}

\begin{document}

\baselineskip=18pt

\begin{center}

{\Large \tit.arg} \if1\blind{\footnote{ \baselineskip=12pt
Corresponding Author:
Rong Zhu (rongzhu@amss.ac.cn).
Zhu's work was partially supported by National Natural Science Foundation of China (Grant nos.  11301514 and 71532013).
Zou's work was partially supported by National Natural Science Foundation of China (Grant nos. 11529101 and 11331011) and the Ministry of Science and Technology of China (Grant no. 2016YFB0502301).\\
AMS 2010 subject classification. 62D05.
}

\bigskip

  Xianpeng Zong$^\dag$, Rong Zhu$^\ddag$ and Guohua Zou$^\dag$

$^\dag${\it School of Mathematical Sciences, Capital Normal University, Beijing, 100048, China}

\medskip

$^\ddag${\it Academy of Mathematics and Systems Science,
  Chinese Academy of Sciences, Beijing 100190, China}\\
rongzhu@amss.ac.cn

}\fi

 \end{center}

 \noindent{\bf Summary.}
\abst.arg

\bigskip

\noindent {KEY WORDS:} \key.arg

\bigskip

\pagenumbering{arabic}
\setcounter{page}{1}

\date{} \renewcommand{\topfraction}{.95} \linespread{1.36}

\section{Introduction}
\label{sec:intro}

The Horvitz-Thompson (HT) estimator proposed by \cite{HT:52} is widely used in survey sampling.
It has also been applied to other fields such as functional data analysis \citep{CJ11} and the treatment effect \citep{R02}.
The HT estimator is an unbiased estimator constructed via inverse probability weighting.
However, when the inclusion probabilities are highly heterogeneous, i.e., inclusion probabilities of some units are relatively tiny,
the variance of HT estimator would become large due to the inverse probability weighting.
In this paper, we propose an improved Horvitz-Thompson (IHT) estimator to address this problem.

Our approach is to use hard-threshold for the first-order inclusion probabilities.
Specifically, we choose an inclusion probability as the threshold first. Those inclusion probabilities smaller than the threshold are then set to equal the threshold, while the others remain unchanged. By this way, we obtain the modified inclusion probabilities. Finally, we construct an estimator based on these modified inclusion probabilities by using the inverse probability weighting.
We call such an estimator as the IHT estimator.
This method looks very easy but is more efficient than the classical HT estimator.
This hard-threshold approach can be explained as a shrinkage method.
Shrinkage is very commonly used in statistics, such as ridge regression \citep{hoerl:70} and high-dimensional statistics \citep{tibshirani:96}.
In this paper, we use it to reduce the negative effect of highly heterogeneous inclusion probabilities.
Similar to other shrinkage methods, our modification process introduces bias while reduces variance much more, so it improves the estimation efficiency.
We will theoretically and numerically show the improvement from using the modified inclusion probabilities.
In addition to the HT estimator, we also extend this strategy to the ratio estimator, and accordingly, the improved ratio estimator is obtained.



The remainder of the paper is organized as follows.  Section
\ref{sec:HT}  introduces the classical HT estimator and shows its drawback.
Section \ref{sec:Im-HT} proposes our modified inclusion probabilities and the corresponding IHT estimator.
We shall also provide the IHT estimator's properties, and theoretically compare it with the HT estimator in this section.
Section \ref{sec:extension} extends our idea to obtain the improved ratio estimator and shows that our modification is efficient.
Section \ref{sec:empirical} presents numerical evidences from simulations and a real data analysis. Section \ref{sec:conclusion} concludes.  Proofs of theoretical results are given in the Appendix.

\section{HT estimator and its drawback}
\label{sec:HT}
Consider a finite population $U=\{U_1, \cdots, U_N\}$ of size $N$, where $U_k$ denotes the $k$th unit.
For simplicity, we write $U=\{1, \cdots, k, \cdots, N\}$.
For each unit $k$, suppose that the value $y_k$ of the target characteristic $Y$ is measured.
Our aim is to estimate the total, $t_y=\sum_Uy_k$, using a sample $s$ of size $n$ which is randomly
drawn from the population $U$.
We implement unequal probability sampling without replacement.
Denote $\{\pi_k\}_{k=1}^N$ as the first-order inclusion probabilities
and $\{\pi_{kl}\}_{k\neq l}$ as the second-order inclusion probabilities.

\cite{HT:52} proposed the HT estimator as follows
\begin{equation}\label{HT-t}
\hat{t}_{\text{HT}}=\sum\limits_{k\in s}\frac{y_k}{\pi_k}.
\end{equation}
The HT estimator $\hat{t}_{\text{HT}}$ is an unbiased estimator of $t_y$ and
its variance is
\begin{equation}\label{HT-var}
V(\hat{t}_{\text{HT}})={\sum}_{U}\frac{\Delta_{kk}}{\pi_k^2}y_k^2+\dsum\frac{\Delta_{kl}}{\pi_k\pi_l}y_ly_k,
\end{equation}
where $\Delta_{kk}=\pi_k-\pi_k^2$ for all $k$ and $\Delta_{kl}=\pi_{kl}-\pi_k\pi_l$ for all $k\neq l$.

From Eqn. (\ref{HT-var}), when the inclusion probabilities are highly imbalanced, i.e., some $\pi_k$'s are very small,
the variance of the HT estimator may be very large.

\section{Improved HT estimator}
\label{sec:Im-HT}

In this section, we improve the HT estimator in the sense of reducing its mean-squared error (MSE).
The resultant estimator is referenced as the IHT estimator.
For doing this, we first propose modified first-order inclusion probabilities, where the hard-threshold method is used
to reduce the effects of those inclusion probabilities with relatively tiny values.
\begin{definition}\label{mod-pi}
Let $\pi_{(1)}\leq\pi_{(2)}\leq\cdots\leq\pi_{(N)}$ be the ordered values of the first-oder inclusion probabilities $\{\pi_1, \pi_2, \cdots, \pi_N\}$. Assume that there exists an integer $K\geq 2$ such that $\pi_{(K)}\leq (K+1)^{-1}.$
We define the modified first-order inclusion probabilities as follows
    \begin{align}
    \pi_k^*=
    \begin{cases}
    \pi_k   &   \pi_k>\pi_{(K)},    \\
    \pi_{(K)}       &   \pi_k\leq \pi_{(K)},
    \end{cases}
    \quad \quad 1\leq k \leq N.
    \end{align}
\end{definition}
From the definition, we partition the finite population into two parts: $U_1=\{k:\pi_k>\pi_{(K)}\}$ with size $N-K$, and $U_2=\{k:\pi_k\leq \pi_{(K)}\}$ with size $K$.
For $U_1$, the first-order inclusion probabilities keep unchanged, while all of first-order inclusion probabilities for $U_2$ are replaced by $\pi_{(K)}$. From this hard-threshold,
we get our modified first-order inclusion probabilities $\{\pi_k^*\}_{k=1}^N$. Obviously, the choice of $K$ is very important.
In Section \ref{sec:chooseK}, we shall provide a simple way to choose $K$.

{\bf Remark on existence of $K$.} The assumption in Definition \ref{mod-pi} is quite weak. If $\pi_{(2)}>1/(2+1)$, then the sampling fraction $f>\frac{1}{3}-\frac{1}{3N}$.
However that situation that $f>\frac{1}{3}$ rarely happens for large population in practical surveys. Thus,
the inequality that $\pi_{(2)}\leq 1/(2+1)$ generally holds.

Instead of the original first-order inclusion probabilities $\{\pi_k\}_{k=1}^N$, we use our defined modified first-order inclusion probabilities $\{\pi_k^*\}_{k=1}^N$ to construct an improved Horvitz-Thompson (IHT) estimator by inverse probability weighting.
\begin{definition}\label{imp-ht}
The IHT estimator is defined as
\begin{equation}\label{MHT-t}
\hat{t}_{\text{IHT}}=\sum\limits_{k\in s}\frac{y_k}{\pi_k^*}.
\end{equation}
\end{definition}
Unlike the unbiased HT estimator, the IHT estimator is biased. However, this modification would lead to much less MSE due to reducing variance.
Note that our modification idea can be easily extended to the Hansen-Hurwitz estimator \citep{HH:43} for sampling with replacement.

\subsection{Properties of the IHT estimator}
\label{sec:property}
In this section, we derive the properties of our IHT estimator. We first provide the expressions of its bias, variance and MSE in Theorem \ref{iht-pro}, where an unbiased estimator of MSE is also presented.  Then we compare the IHT estimator and the HT estimator  in Theorems \ref{lem1} \& \ref{them:ht}.

\begin{theorem}\label{iht-pro}
The bias and variance of the IHT estimator $\hat{t}_{\text{\text{IHT}}}$ are expressed as
\begin{align}\label{bias-mht}
\text{Bias} (\hat{t}_{\text{IHT}})&={\sum}_{U_2}\left(\frac{\pi_k}{\pi_{(K)}}-1\right)y_k,
\end{align}
and
\begin{align}
\text{Var}(\hat{t}_{\text{IHT}})&={\sum}_U\frac{\Delta_{kk}}{\pi_k^{*2}}y_k^2+\dsum\frac{\Delta_{kl}}{\pi_k^*\pi_l^*}y_ky_l,\label{var-mht}
\end{align}
respectively, where $\Delta_{kk}=\pi_k(1-\pi_k),\Delta_{kl}=\pi_{kl}-\pi_k\pi_l$ $(k\neq l)$ as defined before. 
Therefore, its MSE is given by
\begin{equation}\label{MSE-mht}
\text{MSE}(\hat{t}_{\text{IHT}}) =\left[{\sum}_{U_2}\left(\frac{\pi_k}{\pi_{(K)}}-1\right)y_k\right]^2+{\sum}_U\frac{\Delta_{kk}}{\pi_k^{*2}}y_k^2+\dsum\frac{\Delta_{kl}}{\pi_k^*\pi_l^*}y_ky_l.
\end{equation}
An unbiased estimator of the MSE is
\begin{align}\label{mse-mht}
\widehat{\text{MSE}
}(\hat{t}_{\text{IHT}})=&{\sum}_{s_2}\frac{(\pi_k-\pi_{(K)})^2}{\pi_{(K)}^2\pi_k}y_k^2+\dsumss\frac{(\pi_k-\pi_{(K)})(\pi_l-\pi_{(K)})}{\pi_{(K)}^2\pi_{kl}}y_ky_l\\\notag
&+{\sum}_s\frac{\check{\Delta}_{kk}}{\pi_k^{*2}}y_k^2+\dsums\frac{\check{\Delta}_{kl}}{\pi_k^*\pi_l^*}y_ky_l,
\end{align}
where $\displaystyle\check{\Delta}_{kk}=\frac{\Delta_{kk}}{ \pi_{k}},\check{\Delta}_{kl}=\frac{\Delta_{kl}}{\pi_{kl}}$, $s$ is the sample set, and $s_2=s\cap U_2$. \end{theorem}
\noindent \textbf{Proof.} See Appendix \ref{pthe1}.

To derive the properties of the IHT estimator, we need the following regularity conditions:
{\condition \label{c1}$\displaystyle \min_{i\in U}\pi_i\geq \lambda>0,\min_{i,j\in U}\pi_{ij}\geq \lambda^*>0$, and $$\limsup_{N\rightarrow \infty} n\max_{i\neq j\in U}\mid\pi_{ij}-\pi_i\pi_j\mid<\infty.$$}
{\condition \label{c2}$\displaystyle \max_{i\in U} |y_i|\leq C$ with $C$ a positive constant not depending on $N$.}

Condition \ref{c1} is a common condition imposed on the first-order and second-order inclusion probabilities. The same conditions are used in \cite{Breidt2000Local}, where further comments on \ref{c1} are provided.
Condition \ref{c2} is also a common condition. 

\begin{theorem}\label{lem1}
For the classical HT estimator $\hat{{t}}_\text{HT}$ and the IHT estimator $\hat{{t}}_\text{IHT}$, under the conditions \ref{c1}-\ref{c2}, we have
\begin{align*}
\ \text{Bias}(N^{-1}\hat{{t}}_\text{HT})=0,  \quad\quad\quad\quad\text{Bias}(N^{-1}\hat{{t}}_\text{IHT})=O(n^{-1});
\end{align*}
\text{and}
\begin{align*}
\text{MSE}(N^{-1}\hat{{t}}_\text{HT})=O(n^{-1}),\quad  \text{MSE}(N^{-1}\hat{{t}}_\text{IHT})=O(n^{-1}).
\end{align*}
\end{theorem}
\noindent \textbf{Proof.} See Appendix \ref{plem1}.

From Theorem \ref{lem1}, the squared-bias of our IHT estimator is very small compared to its MSE.
Although our IHT estimator produces an extra bias to reduce the variance, the price for this is relatively small.
The following theorem  theoretically compares the efficiency of the two estimators.
\begin{theorem}\label{them:ht}
Under the conditions \ref{c1}-\ref{c2}, we have
\begin{equation}\label{MSE-compare}
\text{MSE}(N^{-1}\hat{t}_{\text{IHT}})\leq \text{MSE}(N^{-1}\hat{t}_{\text{HT}})+o(n^{-1}).
\end{equation}
Especially, for Poisson sampling, we obtain
\begin{equation}\label{MSE-compare-poisson}
\text{MSE}(N^{-1}\hat{t}_{\text{IHT}})\leq \text{MSE}(N^{-1}\hat{t}_{\text{HT}}),
\end{equation}
 where the strict inequality is true if there exist $k\neq l\in U_2$ such that $(\pi_k-\pi_{(K)})y_k\neq(\pi_l-\pi_{(K)})y_l$.

\end{theorem}
\noindent \textbf{Proof.} See Appendix \ref{pthe2}.

Theorem \ref{them:ht} shows that the IHT estimator is asymptotically more efficient than the classical HT estimator, that is, the MSE of IHT estimator is asymptotically not larger than that of the classical HT estimator. For Poisson sampling, the MSE of IHT estimator is uniformly not larger than that of the classical HT estimator.

\subsection{The choice of $K$}
\label{sec:chooseK}
 The efficiency of the IHT estimator replies on the choice of $K$.
 When $K$ becomes larger, we need to modify more inclusion probabilities and this would cause larger bias.
On the other hand, the improvement of the IHT estimator would not be significant if $K$ becomes smaller. Thus, the threshold $K$ provides a control of the variance-and-bias tradeoff.
Theoretically, Condition \ref{c1}  and the condition $\pi_{(K)}\leq {(K+1)^{-1}}$ of Definition \ref{mod-pi}  implies $K/N=O(N^{-1})$, which provides a guide for choosing $K$ from a theoretical view.
In practice, we propose the following algorithm to choose $K$. Following Algorithm 1, $K$ satisfies the condition $\pi_{(K)}\leq {(K+1)^{-1}}$ of Definition \ref{mod-pi}.

\begin{algorithm}\label{alg:K}
  \caption{The choice of $K$}
  \begin{enumerate}[Step (i)]
  \item Obtain the ordered inclusion probabilities $\{\pi_{(1)}, \pi_{(2)}, \cdots, \pi_{(N)}\}$ by sorting $\{\pi_k\}_{k=1}^N$ from small to large. Set $K=0$.
  \item Test and modify.\\
  \ \ For $j=1,\cdots,N$:\\
  \ \ \  \ \ \ if $\pi_{(j)}\leq \frac{1}{j+1}$, then we define the modified first-order inclusion probabilities as
  $$\sbf{\pi}^*=\{\underbrace{\pi_{(j)}, \cdots, \pi_{(j)}}_{j-1}, \pi_{(j)}, \pi_{(j+1)}, \cdots, \pi_{(N)}\},$$  and $K=K+1$; \\
  \ \ \  \ \ \ otherwise, stop.
  \end{enumerate}
\end{algorithm}

\section{Extension to the Ratio Estimator}
\label{sec:extension}

When an auxiliary variable is available, the ratio estimator is usually used  to estimate the population total.
In this section, we extend the IHT estimator to the case of ratio estimation.

\subsection{Improved Ratio Estimator}
Denote by $R$ the ratio between the population totals of $Y$ and $Z$ of two characteristic values, i.e.,
\begin{align}
R=\frac{t_y}{t_z}=\frac{\bar{t}_y}{\bar{t}_z},
\end{align}
where $t_y$ and $t_z$ are the totals of the finite populations $Y$ and $Z$,
$\bar{t}_y$ and $\bar{t}_z$ are their means, respectively.
The classical estimator and our modification estimator of $R$ are given by
 \begin{align}
        \hat{R}=\frac{\hat{\bar{t}}_{y\pi}}{\hat{\bar{t}}_{z\pi}}, \text{ and } \hat{R}^*=\frac{\hat{\bar{t}}^*_{y\pi}}{\hat{\bar{t}}^*_{z\pi}},\label{R-ht}
    \end{align}
where $\hat{\bar{t}}_{y\pi}$ and $\hat{\bar{t}}_{z\pi}$ are the HT estimators of $\bar{t}_y$ and $\bar{t}_z$, respectively, while
$\hat{\bar{t}}^*_{y\pi}$ and $\hat{\bar{t}}^*_{z\pi}$ are the IHT estimators using the modified inclusion probabilities.
Specifically,  $\hat{\bar{t}}_{y\pi}=N^{-1}\sum_s\frac{y_k}{\pi_k}$, $\hat{\bar{t}}_{z\pi}=N^{-1}\sum_s\frac{z_k}{\pi_k}$, $\hat{\bar{t}}^*_{y\pi}=N^{-1}\sum_s\frac{y_k}{\pi_k^*}$, and $\hat{\bar{t}}^*_{z\pi}=N^{-1}\sum_s\frac{z_k}{\pi_k^*}$.

We assume that the population total $t_z$ of $Z$ is known. To estimate the population total $t_y$ of $Y$, the classical ratio estimator is given by
 \begin{align}
        \hat{Y}_{R}=t_z\cdot\frac{\hat{\bar{t}}_{y\pi}}{\hat{\bar{t}}_{z\pi}}.
    \end{align}
Alternatively, our improved ratio estimator of $t_y$  based on the modified inclusion probabilities is expressed as
 \begin{align}
        \hat{Y}^*_{R}=t_z\cdot\frac{\hat{\bar{t}}^*_{y\pi}}{\hat{\bar{t}}^*_{z\pi}}.
    \end{align}

\subsection{Properties of the improved ratio estimator}
To show theoretically that the improved ratio estimator $\hat{Y}_R^*$ is more efficient than the classical ratio estimator $\hat{Y}_R$, we need the following regularity conditions:
{\condition \label{c3}$\displaystyle\lim_{N\rightarrow\infty}\frac{n}{N}=c,$ where $c\in (0,1)$ is a constant.}
{\condition \label{c4}$\displaystyle {\max_{i\neq j\neq k\in U} (\pi_{ijk}-\pi_{ij}\pi_k)=O(n^{-1})},$ and
$$ \max_{i\neq j\neq k\neq l\in U}(\pi_{ijkl}-4\pi_{ijk}\pi_{l}+6\pi_{ij}\pi_{k}\pi_{l}-3\pi_{i}\pi_{j}\pi_{k}\pi_{l})=O(n^{-2}).$$
}
Condition \ref{c3} is a common condition. The same condition is used in \cite{Breidt2000Local}.
Condition \ref{c4} is a mild assumption on the third-order and fourth-order inclusion probabilities. In Appendix \ref{exp}, we present some frequent examples which satisfy Condition \ref{c4}.

Comparing our improved estimators with the classical estimators, we have the following result.
{\theorem\label{them:r} If Conditions \ref{c1}-\ref{c4} are satisfied,  and $c_1\leq z_k\leq c_2$ for all $k\in U$ with $c_1$ and $c_2$ some positive constants, then
    $$\text{MSE}(\hat{R}^*)\leq \text{MSE}(\hat{R})+o(n^{-1}).$$
Furthermore,
    $$\text{MSE}(N^{-1}\hat{Y}^*_R)\leq \text{MSE}(N^{-1}\hat{Y}_R)+o(n^{-1}).$$
}
\noindent \textbf{Proof.} See Appendix \ref{pthe3}.

Like Theorem \ref{them:ht}, Theorem \ref{them:r} shows that the proposed method improves the classical ratio estimators up to order $o(n^{-1})$.

\section{Numerical Studies}
\label{sec:empirical}
In this section, we assess the empirical performance of our IHT estimator by four synthetic examples and one real example. 
 We consider the following two cases: the estimations of population total and population ratio,
where our IHT strategies are compared with the corresponding classical HT methods.
We measure the efficiency improvement in term of $Re=\frac{|\text{MSE}^{\text{HT}}-\text{MSE}^{\text{IHT}}|}{\text{MSE}^{\text{HT}}}\times100\%$,  where MSE$^{\text{HT}}$ and MSE$^{\text{IHT}}$ denote the MSE of the classical HT estimators and IHT estimators, respectively.

\subsection{Simulations}
\label{sec:simu}


\textbf{Example 1: An illustrative example}

We generate a finite population $Y$ of size $N=3000$, where the $k$-th unit value $y_k=|y_{0k}|$ and $y_{0k}\sim N(0,1)$.
 Our aim is to estimate the population mean $\bar{Y}=\frac{1}{N}{\sum}_Uy_k$.
We perform Poisson sampling according to the inclusion probabilities set as follows
\begin{align*}
        \pi_1&=\pi_2\quad=\dots=\pi_{1000}=0.2,\\
        \pi_{1001}&=\pi_{1002}=\dots=\pi_{2000}=0.001,\\
        \pi_{2001}&=\pi_{2002}=\dots=\pi_{3000}=0.08.
\end{align*}
In this example, the HT estimator could be less efficient since one third inclusion probabilities are 0.001, tiny relative to 0.08 or 0.2.
From our hard-threshold strategy, we replace these tiny probabilities with 0.08, so the modified inclusion probabilities are given by
    \begin{align*}
        \pi_1^*&=\pi_2^*\quad=\dots=\pi_{1000}^*=0.2,\\
        \pi_{1001}^*&=\pi_{1002}^*=\dots=\pi_{2000}^*=0.08,\\
        \pi_{2001}^*&=\pi_{2002}^*=\dots=\pi_{3000}^*=0.08.
    \end{align*}
Note that this modified probabilities are not obtained according to Algorithm 1. It is an illustrative example to show that our hard-threshold can bring efficiency improvement.
By setting  the iteration time $M = 2000$,  we get the simulated biases, variances and MSEs of our IHT estimator and the classical HT estimator.
The results are shown in Table \ref{b1}.
    \begin{table}[h]
        \tabcolsep 4mm
        \caption{Performance of Example 1}
        \begin{center}
        \begin{tabular}{r@{.}lr@{.}lr@{.}lr@{.}lr@{.}lr@{.}lr@{.}l}
        \hline
        \multicolumn{2}{c}{MSE$^{\text{HT}}$}&\multicolumn{2}{c}{MSE$^{\text{IHT}}$}&\multicolumn{2}{c}{Bias$^{\text{HT}}$}&\multicolumn{2}{c}{Bias$^{\text{IHT}}$}
        &\multicolumn{2}{c}{Var$^{\text{HT}}$}&\multicolumn{2}{c}{Var$^{\text{IHT}}$}&\multicolumn{2}{c}{$Re \uparrow$ }\\ \hline
        0&1187 &0&0751 &5&3740E-06 &0&0723 &0&1187 &0&0029 &36&71\%  \\\hline
        \end{tabular}
        \end{center}\label{b1}
    \end{table}

It is seen from the table that the variance of the classical HT estimator is much larger than that of the IHT estimator, so it loses its efficiency compared to the IHT estimator  although  the classical HT estimator is unbiased.
Thus, the IHT estimator has much less MSE than that of the classical HT estimator. Specifically, the MSE of the IHT estimator decreases 36.7\%.
Furthermore, in order to show the variations of both estimators, we plot their values in Figure 1. Figure 1 clearly displays that although there is some bias for the IHT estimator, its variation is much less than that of the classical HT estimator.
These observations empirically verify our theoretical results.
    \begin{figure}[h]
        \centering
        \includegraphics[width=0.67\textwidth]{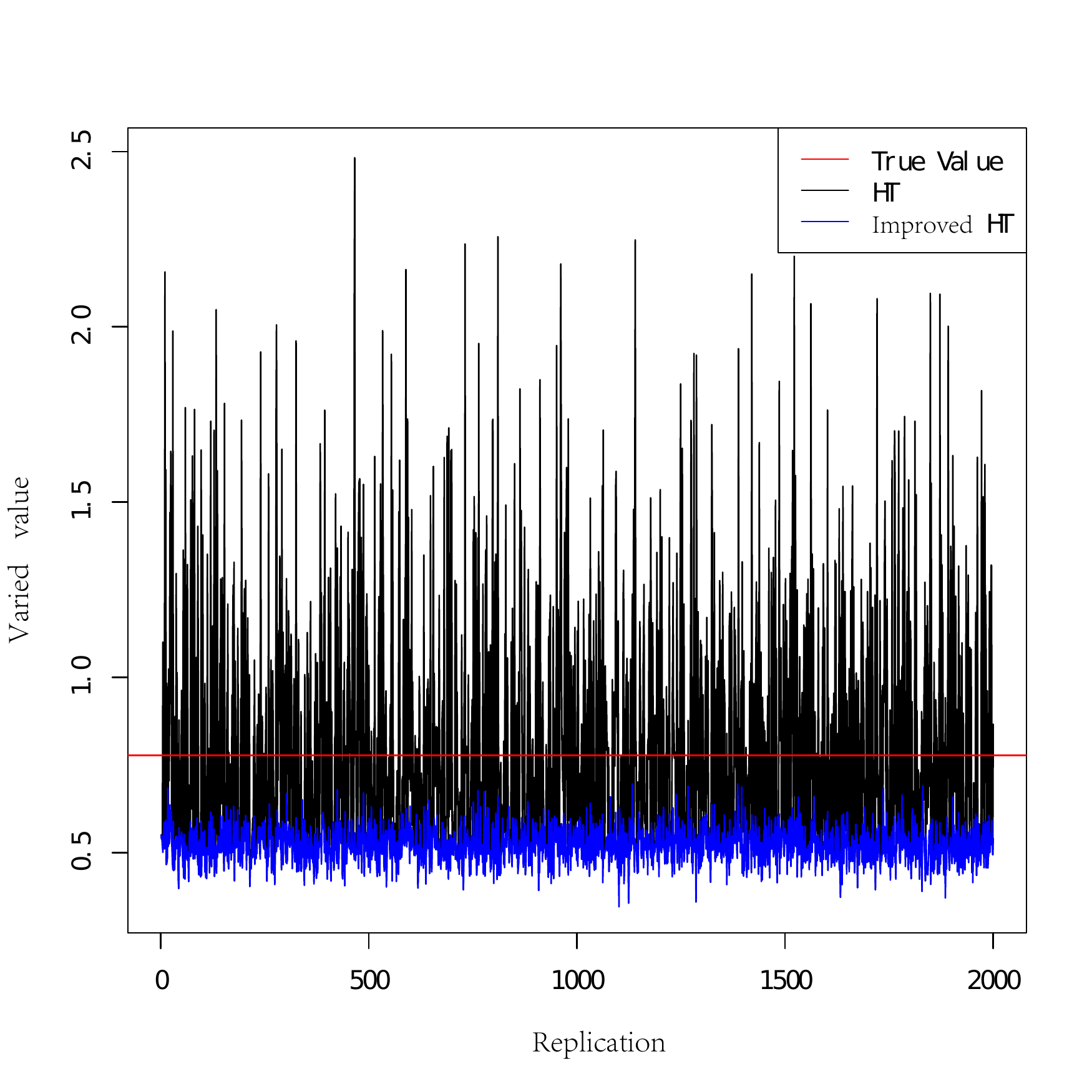}\\
        \caption{The plots of both estimators in Example 1}\label{mn_1}
    \end{figure}

\noindent\textbf{Example 2: $\pi_i$' depending on auxiliary variable}

\begin{figure}[h]
        \centering
        \subfigure[Poisson sampling]{
        \label{pot}
        \includegraphics[width=0.9\textwidth]{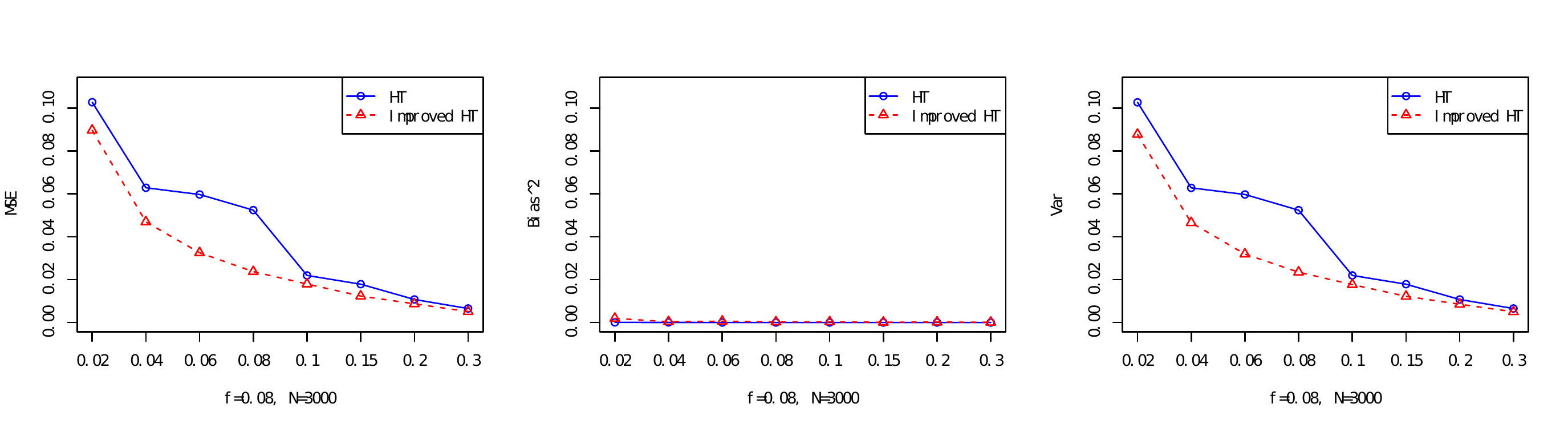}}
        \subfigure[PPS samping]{
        \label{ppst}
        \includegraphics[width=0.9\textwidth]{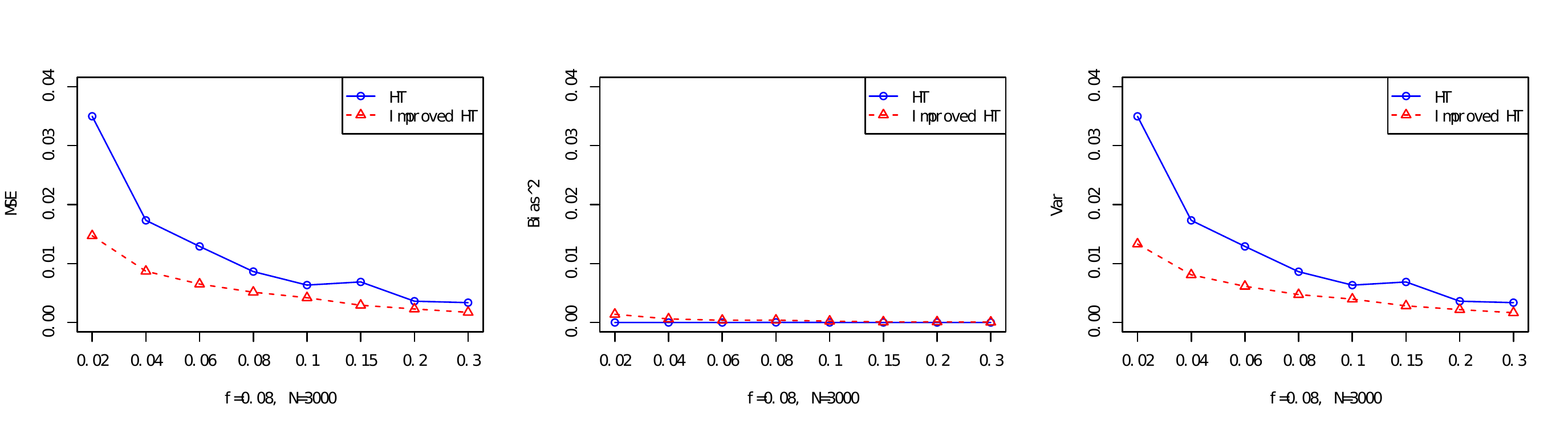}}
        \subfigure[$\pi$PS samping]{
        \label{pipst}
        \includegraphics[width=0.9\textwidth]{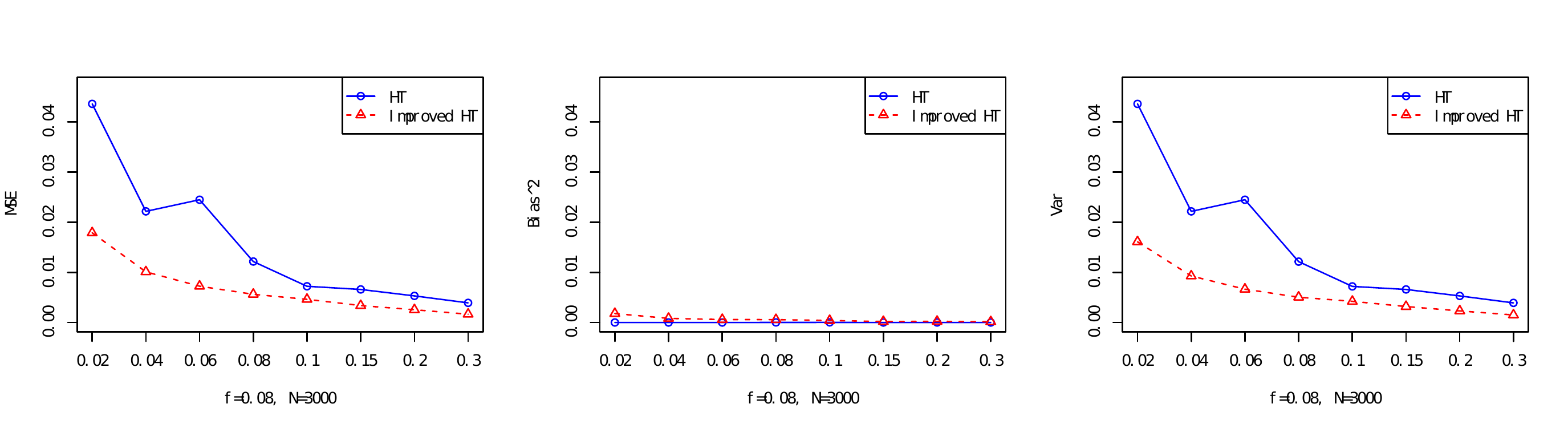}}
        \caption{The performance of our IHT estimator and the classical HT estimator in Example 2, where $\rho=0.8$. From left to right: the MSE performance, the squared-bias performance, and the variance performance.}
        \label{fig:example2}
\end{figure}

 We generate the finite population $Y$ of size $N=3000$ as follows:  $y_k=\sqrt{3}\cdot \rho\cdot x_k+\sqrt{3-3\rho^2}\cdot |e_k|$,
where $x_k$ and $e_k$ are independently generated from $U(0,2)$ and $N(0,1)$ respectively, 
 and $0\leq \rho\leq 1$ controling the correlation of $Y$ and $X$.
We consider three sampling methods: Poisson sampling, PPS sampling and $\pi$PS sampling. The sampling fraction $\displaystyle f=\frac{n}{N}=0. 02,0.04,0.06,0.08,0.10,0.15,0.20,0.30$.
We report the results in Figure \ref{fig:example2}, where $\rho=0.8$. 
From this figure, we get the same observations as Example 1.  It indicates that our IHT estimator outperforms the classical HT estimator.
By Figures \ref{pot}, \ref{ppst}, and \ref{pipst}, $\pi$PS sampling obtains the biggest advantage of our IHT estimator over the classical HT estimator in terms of efficiency and robustness.
We also list some specific $Re$ values of Figure 2 in Table 4, which show that the improvement is generally substantial.

In order to investigate the effect of $\rho$, we also show the results for different $\rho$ values under $\pi$PS sampling in Table \ref{qpipsb}.
It is observed from the table that no matter what values $\rho$ takes, our IHT estimator has uniformly much less MSE than classical HT estimator.
\begin{table}[h]
      \caption{The performance of Example 2}
      \label{qpipsb}
      \centering
      \begin{tabular}{c| c c | c c | c c | c}
        \hline\hline
        $\rho$& MSE$^{\text{HT}}$ & MSE$^{\text{IHT}}$ & Bias$^{\text{HT}}$ & Bias$^{\text{IHT}}$ & Var$^{\text{HT}}$ & Var$^{\text{IHT}}$ & $Re  \uparrow$\\ \hline
         0     & 3.45E-02 & 1.36E-02 & 3.43E-05 & 5.82E-04 & 3.45E-02 & 1.30E-02 & 60.70\% \\
    0.1   & 2.51E-02 & 1.38E-02 & 1.16E-05 & 8.25E-04 & 2.51E-02 & 1.30E-02 & 44.91\% \\
    0.3   & 2.43E-02 & 1.24E-02 & 4.65E-06 & 8.86E-04 & 2.43E-02 & 1.15E-02 & 48.97\% \\
    0.5   & 2.38E-02 & 1.07E-02 & 9.83E-06 & 8.44E-04 & 2.38E-02 & 9.88E-03 & 54.92\% \\
    0.8   & 9.38E-03 & 5.22E-03 & 3.04E-07 & 3.16E-04 & 9.38E-03 & 4.91E-03 & 44.33\% \\
    0.9   & 4.75E-03 & 2.65E-03 & 7.98E-06 & 2.64E-04 & 4.74E-03 & 2.38E-03 & 44.27\% \\
        \hline
        \end{tabular}
\end{table}

\noindent\textbf{Example 3: $\pi_i$' independent of auxiliary variable}

In this example,  we consider a sampling process which is independent of $X$.
We generate a finite population as in Example 2,  and set the inclusion probabilities $\pi_i \propto |c_i|$, where $c_i\sim N(50,\sigma^2)$. Table \ref{difv} shows the $Re$ values for different $\sigma^2$ ($\sigma^2=5,8,10,15,20,25$) under $\pi$PS sampling, where $\sigma^2$ controls the heterogeneity of $\pi_i$'s.
When $\sigma^2$ becomes larger,  inclusion probabilities become more heterogeneous.
From the table, the $Re$ value increases as the $\sigma^2$ increases. It makes sense since more $\pi_i$'s are modified when $\sigma^2$ becomes larger.

    \begin{table}
      \caption{$Re$ values for different variances in Example 3}\label{difv}
      \centering
      \begin{tabular}{c|cccccccc}
        \hline\hline
        $f$	& 0.02 	& 0.04	& 0.06 	& 0.08	& 0.10	& 0.15	& 0.20	& 0.30\\ \hline
        $\sigma^2$= 5 & 0.35\% & 0.22\% & 0.14\% & 0.14\% & 0.05\% & 0.03\% & 0.01\% & 0.01\% \\
        $\sigma^2$= 8  & 1.24\% & 0.66\% & 0.56\% & 0.72\% & 0.35\% & 0.30\% & 0.00\% & 0.10\% \\
        $\sigma^2$=10 & 3.73\% & 2.47\% & 2.83\% & 0.44\% & 1.48\% & 1.03\% & 1.52\% & 1.06\% \\
        $\sigma^2$=15 & 19.32\% & 13.17\% & 10.77\% & 9.88\% & 12.61\% & 8.76\% & 6.67\% & 6.10\% \\
       $\sigma^2$=20 & 42.27\% & 43.44\% & 44.83\% & 34.42\% & 34.87\% & 34.02\% & 34.36\% & 32.70\% \\
       $\sigma^2$=25 & 59.39\% & 44.25\% & 58.67\% & 59.46\% & 66.56\% & 52.25\% & 48.23\% & 44.04\% \\
        \hline
        \end{tabular}
    \end{table}

\noindent\textbf{Example 4: The estimation of population ratio}

\begin{figure}[h]
        \centering
        \subfigure[$\pi$PS sampling ($\rho_1=0.3,\rho_2=0.4$)]{
        \label{rqpipst}
        \includegraphics[width=0.8\textwidth]{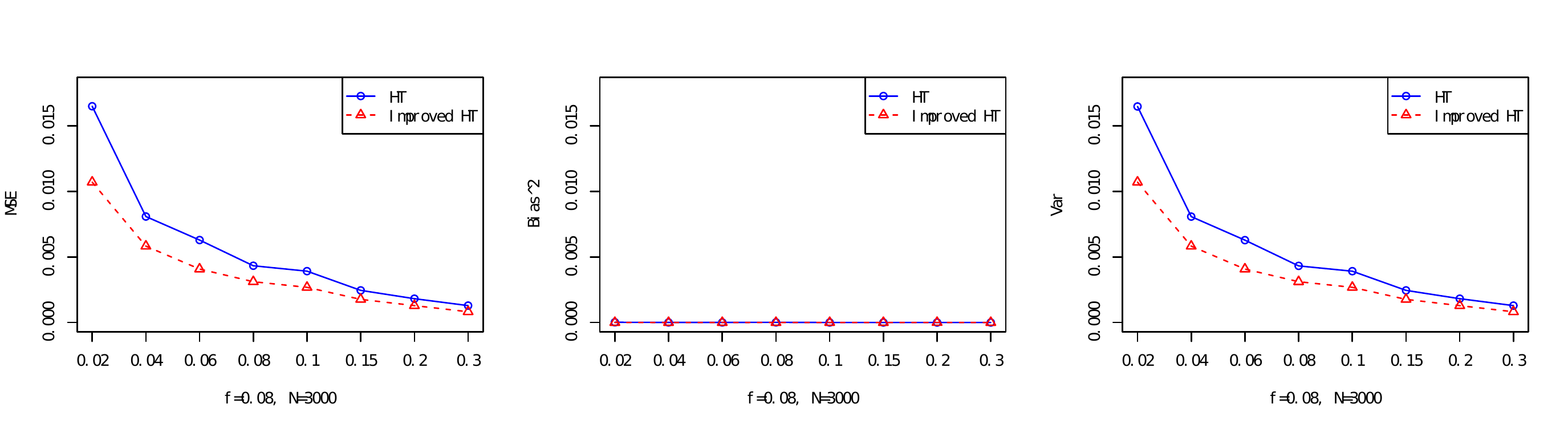}}
        \subfigure[$\pi$PS sampling($\rho_1=0.7,\rho_2=0.8$)]{
        \label{rpipst1}
        \includegraphics[width=0.8\textwidth]{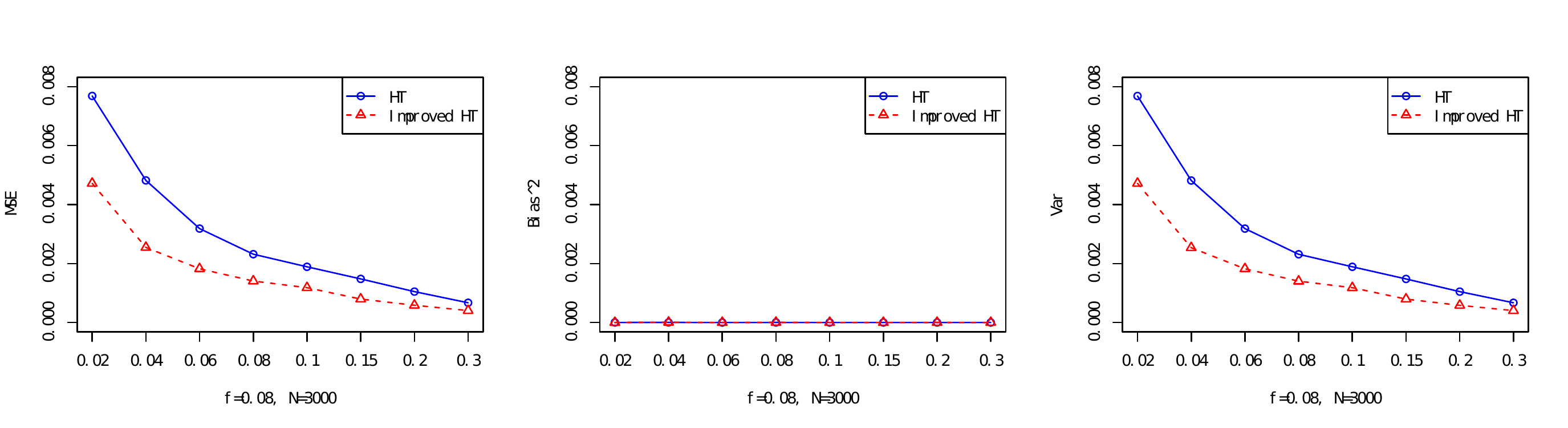}}
        \caption{Performance of Example 4. From left to right: the MSE performance, the squared-bias performance, and the variance performance.}
        \label{fig:example4}
    \end{figure}

We generate two populations $Y$ and $Z$ of size $N=3000$: $y_k=\sqrt{12}\cdot \rho_1\cdot x_k+\sqrt{3-3\rho_1^2}\cdot |e_1|$, and $z_k=\sqrt{12}\cdot \rho_2\cdot x_k+\sqrt{3-3\rho_2^2}\cdot |e_2|$, where auxiliary variable  $x_k\sim U(0,1)$, $e_1\sim N(0,1)$ and $e_2\sim N(0,1)$.
Our aim is to estimate the ratio $\displaystyle R=\frac{t_y}{t_z}$, where $t_y=\sum\limits_{k=1}^Ny_k$ and $t_z=\sum\limits_{k=1}^Nz_k$.
We set $(\rho_1, \rho_2)=(0.3, 0.4)$ or $(0.7,0.8)$, and report the results in Figure \ref{fig:example4}.
From Figures \ref{rqpipst} and \ref{rpipst1}, similar to the estimation of population total in examples above, our improved estimator outperforms the classical estimator.  We also list some specific $Re$ values of Figure 3 in Table 4, which show that the MSEs decrease above 25\%. 

    \begin{table}[h]
      \caption{Some specific $Re$ values of Figures \ref{fig:example2} \& \ref{fig:example4} in Examples 2 \& 4}\label{figre}
      \centering
      \begin{tabular}{c|cccccccc}
        \hline\hline
        $f$ 	& 0.02 	& 0.04	& 0.06 	& 0.08	& 0.10	& 0.15	& 0.20	& 0.30\\ \hline
    Figure \ref{pot} & 12.73\% & 25.33\% & 45.52\% & 54.71\% & 18.15\% & 30.94\% & 18.96\% & 21.99\% \\
    Figure \ref{ppst}  & 57.92\% & 49.78\% & 49.48\% & 40.52\% & 33.81\% & 57.44\% & 36.45\% & 48.70\% \\
   Figure \ref{pipst} & 58.98\% & 54.41\% & 70.42\% & 53.75\% & 36.05\% & 48.72\% & 52.05\% & 57.65\% \\
   \hline
    Figure \ref{rqpipst} & 35.09\% & 27.92\% & 35.16\% & 28.09\% & 31.50\% & 28.00\% & 29.07\% & 36.31\% \\
    Figure \ref{rpipst1} & 38.57\% & 47.18\% & 42.76\% & 39.27\% & 37.49\% & 46.20\% & 44.14\% & 39.55\% \\
        \hline
        \end{tabular}
    \end{table}

\subsection{Real Example}
 We investigate the data set ``Lucy" in R software. This data set includes the variables of  2396 firms:
\emph{ID}, \emph{Level}, \emph{Income}, \emph{Employees}, and \emph{Taxes}.
We set the \emph{Income} as the size of the firm to estimate the \emph{Employees} mean $\bar{Y}$ of the 2300 small or mid-sized firms ($\bar{Y}=60.59$).
We perform $\pi$PS sampling. The sample size $n$ is set among $\{46, 92, 138, 184, 230, 345, 460,690\}$.
We list the results in Table \ref{lucyb}, where the bias, variance, MSE and $Re$ values are reported. We also present the number $K$ chosen by Algorithm 1.
From Table \ref{lucyb}, our IHT estimator has better performance than the classical HT estimator.
As the sampling fraction $f$ increases, the number $K$ decreases. It means that the number of the modified inclusion probabilities decreases as the sampling fraction increases.
This makes sense since the effect of the small inclusion probabilities becomes weak when the sample size increases.
 \begin{table}[h]
      \caption{The performance of estimation for the real data set ``Lucy"}\label{lucyb}
      \centering
      \begin{tabular}{c|cccccccc}
        \hline\hline
        $n$&{46}&{92}&{138}&{184}&{230}&{345}&{460}&{690}  \\ \hline
       {MSE$^{\text{HT}}$} & 42.60 & 20.80 & 26.87 & 9.30  & 6.97  & 8.01  & 6.40  & 2.99 \\
        {MSE$^{\text{IHT}}$}& 28.27 & 14.05 & 10.18 & 7.75  & 5.70  & 3.77  & 2.85  & 1.76 \\\hline
        {Bias$^{\text{HT}}$}& 0.0092 & 0.0002 & 0.0004 & 0.0020 & 0.0041 & 0.0001 & 0.0005 & 0.0112 \\
        {Bias$^{\text{IHT}}$} & 0.7520 & 0.3375 & 0.2562 & 0.1093 & 0.1253 & 0.0831 & 0.0539 & 0.0626 \\\hline
        {Var$^{\text{HT}}$} & 42.59 & 20.80 & 26.87 & 9.30  & 6.97  & 8.01  & 6.40  & 2.97 \\
        {Var$^{\text{IHT}}$} & 27.52 & 13.71 & 9.92  & 7.64  & 5.57  & 3.68  & 2.79  & 1.70 \\\hline
        ${Re} \uparrow$  & 33.64\% & 32.46\% & 62.13\% & 16.75\% & 18.31\% & 53.01\% & 55.49\% & 41.09\% \\\hline
        {$K$}	  & 166   & 100   & 72    & 59    & 49    & 36    & 29    & 21 \\
        \hline
        \end{tabular}
    \end{table}

\section{Concluding Remarks}\label{sec:conclusion}
In this paper, we have proposed a novel and simple method to improve the Horvitz-Thompson estimator in survey sampling.
Compared with the classical HT estimator, the proposed IHT estimator improves the estimation accuracy at the expense of introducing small bias.
Empirical studies show that the improvement can be substantial. The new idea has also been  used to construct the improved ratio estimator. Naturally, applying the new method to the regression estimation problem is of interest as well, and this warrants our further study.

The choice of the threshold $K$ is important in our method. Although we have suggested an easy algorithm for choosing $K$, it is may not be optimal. How to get the most efficient of way choosing $K$ is a meaningful topic for the future research.

\renewcommand\refname{References}
\baselineskip=16pt
\bibliographystyle{Chicago} 
\bibliography{SHT}

\renewcommand{\thesection}{A.\arabic{section}}
\renewcommand{\thesubsection}{A.\arabic{subsection}}
\section*{Appendix}

\subsection{Proof of Theorem \ref{iht-pro}}\label{pthe1}

To obtain the MSE of the IHT estimator, we first define $I_k=1$ or $0$, $k=1,\cdots, N$, if the $k$th unit is drawn or not, then
\begin{center}
$E(I_k)=\pi_k, Var(I_k)=\Delta_{kk}$, $Cov(I_k,I_l)=\Delta_{kl}$ for $k\neq l$,
\end{center}
where $\Delta_{kk}=\pi_k(1-\pi_k),\Delta_{kl}=\pi_{kl}-\pi_k\pi_l.$
So the bias of the IHT estimator is
\begin{align}\label{dev}
        Bias(\hat{t}_{\text{IHT}})
        =E\left({\sum}_U\frac{y_k}{\pi_k^*}I_k\right)-{\sum}_Uy_k
        ={\sum}_{U_2}\left(\frac{\pi_k}{a}-1\right)y_k,
    \end{align}
    where $a=\pi_{(K)}$. It follows that
    \begin{align}
        Bias^2(\hat{t}_{\text{IHT}})&=\left[{\sum}_{U_2}\left(\frac{\pi_k}{a}-1\right)y_k\right]^2.\label{z2}
    \end{align}
    The variance of the IHT estimator is given by
    \begin{align}
        Var(\hat{t}_{\text{IHT}})&=Var\left({\sum}_s{y_k}/{\pi_k^*}\right)=Var\left({\sum}_U\frac{y_k}{\pi_k^*}I_k\right)\nonumber\\
        &={\sum}_U\left[\left(\frac{y_k}{\pi_k^*}\right)^2Var(I_k)\right]+\dsum\left(\frac{y_k}{\pi_k^*}\frac{y_l}{\pi_l^*}Cov(I_k,I_l)\right)\nonumber\\
        &={\sum}_{U_1}\frac{\Delta_{kk}}{\pi_k^{2}}y_k^2+{\sum}_{U_2}\frac{\Delta_{kk}}{a^2}y_k^2+\dsum\frac{\Delta_{kl}}{\pi_k^*\pi_l^*}y_ky_l.\label{z3}
    \end{align}
    Combining (\ref{z2}) and (\ref{z3}), we obtain
    \begin{align}\label{z4}
        MSE(\hat{t}_{\text{IHT}})=&Bias^2(\hat{t}_{\text{IHT}})+Var(\hat{t}_{\text{IHT}})\\
        =&\left[{\sum}_{U_2}\left(\frac{\pi_k}{a}-1\right)y_k\right]^2+{\sum}_U\frac{\Delta_{kk}}{\pi_k^{*2}}y_k^2+\dsum\frac{\Delta_{kl}}{\pi_k^*\pi_l^*}y_ky_l\nonumber\\
        =&\left\{{\sum}_{U}\frac{\Delta_{kk}}{\pi_k^{2}}y_k^2+\left[{\sum}_{U_2}\left(\frac{\pi_k}{a}-1\right)y_k\right]^2\right\}+{\dsum\frac{\Delta_{kl}}{\pi_k^*\pi_l^*}y_ky_l}\nonumber\\
        \triangleq& F_1+F_2.\nonumber
    \end{align}

For the MSE estimator $\widehat{MSE}(\hat{t}_{\text{IHT}})$ in Eqn. (\ref{mse-mht}), we have $E(\widehat{MSE}(\hat{t}_{\text{IHT}}))= MSE(\hat{t}_{\text{IHT}})$.
Therefore, Theorem 1 is proved.
     \begin{flushright}
    $\Box$
    \end{flushright}

\subsection{Proof of Theorem \ref{lem1}}\label{plem1}

 Using the conditions \ref{c1} and \ref{c2}, we see that
 $\lambda\leq\pi_k\leq a\leq1$ for each $k \in U_2$, and $\displaystyle\max_{k\neq l\in U_2}\mid\pi_{kl}-\pi_k\pi_l\mid=O(n^{-1})$. Then, from Eqn. (\ref{HT-var}), we have
 \begin{align*}
 \mid E(\hat{\bar{t}}_{HT}-\bar{t})^2\mid
 &=\left| \frac{1}{N^2}\sum\limits_{U}\frac{\Delta_{kk}}{\pi_k^2}y_k^2+ \frac{1}{N^2}\dsum\frac{\Delta_{kl}}{\pi_k\pi_l}y_ly_k\right|\\
 &\leq \frac{1}{N^2}\sum\limits_{U}\frac{1-\pi_k}{\pi_k}y_k^2+ \frac{1}{N^2}\dsum\left|\frac{\pi_{kl}-\pi_k\pi_l}{\pi_k\pi_l}\right|| y_l y_k|\\
 &=O(n^{-1}).
 \end{align*}
Similarly, by the MSE of the IHT estimator given in (\ref{MSE-mht}), we observe
 \begin{align*}
\mid E(\hat{\bar{t}}_{\text{IHT}}-\bar{t})^2\mid
 &=\left|\left[\frac{1}{N}{\sum}_{U_2}\left(\frac{\pi_k}{a}-1\right)y_k\right]^2+\frac{1}{N^2}{\sum}_U\frac{\Delta_{kk}}{\pi_k^{*2}}y_k^2+\frac{1}{N^2}\dsum\frac{\Delta_{kl}}{\pi_k^*\pi_l^*}y_ky_l\right|\\
 &\leq\left[\frac{K}{N} \frac{1}{K}{\sum}_{U_2}\left(\frac{\pi_k}{a}-1\right)y_k\right]^2+\frac{1}{N^2}{\sum}_U\left|\frac{\pi_k(1-\pi_k)}{\pi_k^{*2}}\right| y_k^2\\
 &+ \frac{1}{N^2}\dsum\left|\frac{\pi_{kl}-\pi_k\pi_l}{\pi_k^*\pi_l^*}\right|| y_ky_l|\\
 &=O(n^{-1}).
 \end{align*}
From Eqn. (\ref{dev}), and the conditions \ref{c1} and \ref{c2}, it is readily seen that
\begin{align*}
      Bias(\hat{\bar{t}}_{\text{IHT}})&=\left|\frac{1}{N}{\sum}_{U_2}\left(\frac{\pi_k}{a}-1\right)y_k\right|\leq\frac{K}{N}\frac{1}{K}{\sum}_{U_2}\left|\frac{\pi_k}{a}-1\right||y_k|\leq \frac{K}{N}\frac{1}{K}{\sum}_{U_2}|y_k|=O(n^{-1}),
\end{align*}
where the third and fourth steps are valid due to  $\lambda\leq\pi_k\leq a\leq1$ for each $k \in U_2$ and $K/N=O(n^{-1})$, respectively.
     \begin{flushright}
    $\Box$
    \end{flushright}
\subsection{Proof of Theorem \ref{them:ht}}\label{pthe2}

From Eqn. (\ref{HT-var}), since the classical HT estimator is unbiased, we have
\begin{align}
        MSE(\hat{Y}_{HT})
        =\left\{{\sum}_{U_1}\frac{\Delta_{kk}}{\pi_k^2}y_k^2+{\sum}_{U_2}\frac{\Delta_{kk}}{\pi_k^2}y_k^2\right\}+{\dsum\frac{\Delta_{kl}}{\pi_k\pi_l}y_ky_l}\triangleq F_3+F_4.\label{z1}
\end{align}

To illustrate the effectiveness of the new estimator, we compare  Eqn. (\ref{z4}) and  Eqn. (\ref{z1}).\\
We prove $F_1 \geq F_3$ first.
It is clear that
    \begin{align*}
        F_3-F_1&={\sum}_{U_1}\frac{\Delta_{kk}}{\pi_k^2}y_k^2+{\sum}_{U_2}\frac{\Delta_{kk}}{\pi_k^2}y_k^2    -\left\{{\sum}_{U_1}\frac{\Delta_{kk}}{\pi_k^{2}}y_k^2+{\sum}_{U_2}\frac{\Delta_{kk}}{a^2}y_k^2+\left[{\sum}_{U_2}\left(\frac{\pi_k}{a}-1\right)y_k\right]^2\right\}\\
        &={\sum}_{U_2}\frac{\Delta_{kk}}{\pi_k^2}y_k^2-{\sum}_{U_2}\frac{\Delta_{kk}}{a^2}y_k^2-\left[{\sum}_{U_2}\left(\frac{\pi_k}{a}-1\right)y_k\right]^2\\
        &={\sum}_{U_2}\frac{(a^2-\pi_k^2)(1-\pi_k)}{a^2\pi_k}y_k^2-\left[{\sum}_{U_2}\left(\frac{\pi_k}{a}-1\right)y_k\right]^2\\
        &\triangleq D-C
    \end{align*}
   Using Cauchy-Schwarz inequality, we have
    \begin{align}
        C=\left({\sum}_{U_2}\frac{\pi_k-a}{a}y_k\right)^2\leq K{\sum}_{U_2}\frac{(\pi_k-a)^2}{a^2}y_k^2\triangleq{E},\label{z5}
    \end{align}
    where the strict inequality holds if there exist $k\neq l\in U_2$ such that $(\pi_k-\pi_{(K)})y_k\neq(\pi_l-\pi_{(K)})y_l$.
    Further,
    \begin{align*}
        D-E&={\sum}_{U_2}\frac{(a^2-\pi_k^2)(1-\pi_k)}{a^2\pi_k}y_k^2-K{\sum}_{U_2}\frac{(\pi_k-a)^2}{a^2}y_k^2\\
        &={\sum}_{U_2}\frac{(a-\pi_k)\big[(1-\pi_k-K\pi_k)a+(\pi_k-\pi_k^2+K\pi_k^2)\big]}{a^2\pi_k}y_k^2.
    \end{align*}
    From Definition \ref{mod-pi}, we have  $\pi_k \leq a \leq {(K+1)}^{-1}$ for each $k \in U_2$, thus $D-E \geq 0.$ So $F_3-F_1=D-C\geq D-E\geq 0$ holds.

    As a special case, for Poisson sampling, we have $F_4=F_2=0$. Hence, we obtain
    \begin{equation*}
        \text{MSE}(N^{-1}\hat{t}_{\text{\text{IHT}}})\leq \text{MSE}(N^{-1}\hat{t}_{\text{HT}}).
    \end{equation*}

    For the terms $F_2$ and $F_4$, we note that
     \begin{align*}
        F_2-F_4=&\dsum\frac{\Delta_{kl}}{\pi_k^*\pi_l^*}y_ky_l-\dsum\frac{\Delta_{kl}}{\pi_k\pi_l}y_ky_l\\
        =&
        {\dsumuu\left(\frac{\Delta_{kl}}{a^2}-\frac{\Delta_{kl}}{\pi_k\pi_l}\right)y_ky_l}
        +{\sum_{k\in U_1}\sum_{l\in U_2}\left(\frac{\Delta_{kl}}{a\pi_k}-\frac{\Delta_{kl}}{\pi_k\pi_l}\right)y_ky_l}
        +{\sum_{k\in U_2}\sum_{l\in U_1}\left(\frac{\Delta_{kl}}{a\pi_l}-\frac{\Delta_{kl}}{\pi_k\pi_l}\right)y_ky_l}\\
        \triangleq&{\Delta_1}+{\Delta_2}+{\Delta_3}.
    \end{align*}
    Using the conditions \ref{c1} and \ref{c2}, it is seen that
    \begin{align*}
     \frac{\mid\Delta_1\mid}{N^2}&=\frac{1}{N^2}\left|\dsumuu\frac{\pi_{kl}-\pi_k\pi_l}{\pi_k\pi_l}\left(\frac{\pi_k\pi_l}{a^2}-1\right)y_ky_l\right|\leq \frac{1}{N^2}\dsumuu\left|\frac{\pi_{kl}-\pi_k\pi_l}{\pi_k\pi_l}\right|\left| \frac{\pi_k\pi_l}{a^2}-1\right|| y_ky_l|\\
     &\leq \frac{K^2}{N^2}\frac{1}{K^2}\dsumuu\left|\frac{\pi_{kl}-\pi_k\pi_l}{\pi_k\pi_l}\right|| y_ky_l|
     =O(n^{-3}),
    \end{align*}
    where the third and fourth steps are valid due to  $\lambda\leq\pi_k\leq a\leq1$ for each $k \in U_2$, $K/N=O(n^{-1})$, and $\displaystyle\max_{k\neq l\in U_2}\mid\pi_{kl}-\pi_k\pi_l\mid=O(n^{-1})$.
    Similarly, we obtain
    \begin{align*}
     \frac{\mid\Delta_2\mid}{N^2}&=\frac{1}{N^2}\left|\sum_{k\in U_1}\sum_{l\in U_2}\frac{\pi_{kl}-\pi_k\pi_l}{\pi_k\pi_l}\left(\frac{\pi_l}{a}-1\right)y_ky_l\right|
     \leq \frac{1}{N^2}\sum_{k\in U_1}\sum_{l\in U_2}\left|\frac{\pi_{kl}-\pi_k\pi_l}{\pi_k\pi_l}\right|\left|\frac{\pi_l}{a}-1\right|| y_ky_l| \\
     &\leq \frac{1}{N^2}\sum_{k\in U_1}\sum_{l\in U_2}\mid\frac{\pi_{kl}-\pi_k\pi_l}{\pi_k\pi_l}\mid \mid  y_ky_l\mid
     =O(n^{-2}),
    \end{align*}
   and $\displaystyle\frac{\mid\Delta_3\mid}{N^2}=O(n^{-2})$.

Thus, together with $F_3\geq F_1$, we have $\displaystyle F_3+F_4+o\left(\frac{N^2}{n}\right)\geq F_1+F_2,$ that is,
    \begin{equation*}
    \text{MSE}(N^{-1}\hat{t}_{\text{\text{IHT}}})\leq \text{MSE}(N^{-1}\hat{t}_{\text{HT}})+o(n^{-1}).
    \end{equation*}
    \begin{flushright}
    $\Box$
    \end{flushright}

\subsection{Proof of Theorem \ref{them:r}}\label{pthe3}

First note that
$$(\hat{R}-R)^2=\left(\frac{\hat{\bar{t}}_{y\pi}-R\hat{\bar{t}}_{z\pi}}{\hat{\bar{t}}_{z\pi}}\right)^2
=\frac{(\hat{\bar{t}}_{y\pi}-R\hat{\bar{t}}_{z\pi})^2}{\bar{t}_z^2}-
\frac{(\hat{\bar{t}}_{z\pi}^2-\bar{t}_z^2)(\hat{\bar{t}}_{y\pi}-R\hat{\bar{t}}_{z\pi})^2}{\bar{t}_z^2\ \hat{\bar{t}}_{z\pi}^2}\triangleq{\Rma{1}}+{\Rma{3}},$$
and
$$(\hat{R}^*-R)^2=\left(\frac{\hat{\bar{t}}_{y\pi}^*-R\hat{\bar{t}}_{z\pi}}{\hat{\bar{t}}_{z\pi}^*}\right)^2
=\frac{(\hat{\bar{t}}_{y\pi}^*-R\hat{\bar{t}}_{z\pi}^*)^2}{\bar{t}_z^2}-
\frac{(\hat{\bar{t}}_{z\pi}^{*2}-\bar{t}_z^2)(\hat{\bar{t}}_{y\pi}^*-R\hat{\bar{t}}_{z\pi}^*)^2}{\bar{t}_z^2\ \hat{\bar{t}}_{z\pi}^{*2}}\triangleq{\Rma{2}}+{\Rma{4}}.$$
Let $u_k=y_k-Rz_k.$ By Theorem \ref{them:ht}, we have
\begin{align*}
   E(\hat{\bar{t}}_u^*-\bar{t}_u)^2\leq E(\hat{\bar{t}}_u-\bar{t}_u)^2+o(n^{-1}).
\end{align*}
Thus, for the terms \Rma{1} and \Rma{2}, we get
\begin{align}\label{rp}
E(\Rma{1})\leq E(\Rma{2})+o(n^{-1}).
\end{align}
Now, we need to prove that the expectations of \Rma{3} and \Rma{4} are  negligible.
Observe that,
\begin{align*}
\mid E(\Rma{3})\mid&=\left| E\frac{(\hat{\bar{t}}_{z\pi}+\bar{t}_z)(\hat{\bar{t}}_{z\pi}-\bar{t}_z)(\hat{\bar{t}}_{y\pi}-R\hat{\bar{t}}_{z\pi})^2}{\bar{t}_z^2\ \hat{\bar{t}}_{z\pi}^2}\right|\\
&\leq E\frac{\mid\hat{\bar{t}}_{z\pi}+\bar{t}_z\mid\mid\hat{\bar{t}}_{z\pi}-\bar{t}_z\mid(\hat{\bar{t}}_{y\pi}-R\hat{\bar{t}}_{z\pi})^2}{\bar{t}_z^2\ \hat{\bar{t}}_{z\pi}^2}\\
&\leq \frac{ Z^*+\mid\bar{t}_z\mid}{\bar{t}_z^2\ Z_*^2} E\left(\mid\hat{\bar{t}}_{z\pi}-\bar{t}_z\mid(\hat{\bar{t}}_{y\pi}-R\hat{\bar{t}}_{z\pi})^2\right)\\
&\leq \frac{ Z^*+\mid\bar{t}_z\mid}{\bar{t}_z^2\ Z_*^2} \sqrt{E(\hat{\bar{t}}_{z\pi}-\bar{t}_z) ^2 E(\hat{\bar{t}}_{y\pi}-R\hat{\bar{t}}_{z\pi})^4},
\end{align*}
where $\displaystyle Z^*=\frac{n}{N}\max_{k\in U}\left(\frac{z_k}{\pi_k}\right), Z_*=\frac{n}{N}\min_{k\in U}\left(\frac{z_k}{\pi_k}\right)$.
Similarly, $$\mid E(\Rma{4})\mid\leq\frac{ \tilde{Z}^*+\mid\bar{t}_z\mid}{\bar{t}_z^2\ \tilde{Z}_*^2} \sqrt{E(\hat{\bar{t}}_{z\pi}^*-\bar{t}_z) ^2 E(\hat{\bar{t}}_{y\pi}^*-R\hat{\bar{t}}_{z\pi}^*)^4},$$
 where $\displaystyle \tilde{Z}^*=\frac{n}{N}\max_{k\in U}\left(\frac{z_k}{\pi_k^*}\right), \tilde{Z}_*=\frac{n}{N}\min_{k\in U}\left(\frac{z_k}{\pi_k^*}\right)$.

Using Theorem \ref{lem1} and Lemma \ref{lem2}, we see that $\mid E(\Rma{3})\mid=O(n^{-3/2})$ and $\mid E(\Rma{4})\mid=O(n^{-3/2}).$
Combining these and Eqn. (\ref{rp}), we get
    $$\text{MSE}(\hat{R}^*)\leq \text{MSE}(\hat{R})+o(n^{-1}),$$
which implies
    $$\text{MSE}(N^{-1}\hat{Y}^*_R)\leq \text{MSE}(N^{-1}\hat{Y}_R)+o(n^{-1}).$$
\begin{flushright}
    $\Box$
\end{flushright}
\newpage
\subsection{Discussion on Condition \ref{c4}}\label{exp}
\textbf{Example 1: Simple random sampling without replacement}

Under the simple random sampling without replacement, we have
\begin{align*}
\pi_i=&\frac{n}{N} \quad\quad\quad\quad\quad\quad\quad\quad\quad\quad\quad\  i=1,\dots,N; \\
\pi_{ij}=&\frac{n(n-1)}{N(N-1)};\quad\quad\quad\quad\quad\quad\quad\  i\neq j=1,\dots,N; \\
\pi_{ijk}=&\frac{n(n-1)(n-2)}{N(N-1)(N-2)};\quad\quad\quad\quad i\neq j\neq k=1,\dots,N; \\
\pi_{ijkl}=&\frac{n(n-1)(n-2)(n-3)}{N(N-1)(N-2)(N-3)};\ \ i\neq j\neq k\neq l=1,\dots,N.
\end{align*}
It follows that
$$\pi_{ijk}-\pi_{ij}\pi_k=-\frac{2n(n-1)(N-n)}{N^2(N-1)(N-2)}=O(n^{-1}),$$
where the last equality is from Condition \ref{c3}.
Further, we obtain
\begin{align*}
&\pi_{ijkl}-4\pi_{ijk}\pi_{l}+6\pi_{ij}\pi_{k}\pi_{l}-3\pi_{i}\pi_{j}\pi_{k}\pi_{l}\\
=&(\pi_{ijkl}-\pi_{ijk}\pi_{l})-3(\pi_{ijk}\pi_{l}-\pi_{ij}\pi_{k}\pi_{l})+3(\pi_{ij}\pi_{k}\pi_{l}-\pi_{i}\pi_{j}\pi_{k}\pi_{l})\\
=&3\frac{n(n-1)(n-2)(n-N)}{N^2(N-1)(N-2)(N-3)}-6\frac{n^2(n-1)(n-N)}{N^3(N-1)(N-2)}+3\frac{n^3(n-N)}{N^4(N-1)}\\
=&3\frac{n(n-1)(n-N)(3n-2N)}{N^3(N-1)(N-2)(N-3)}+\frac{n^2(n-N)(N-2n)}{N^4(N-1)(N-2)}\\
=&O(n^{-2}),
\end{align*}
where the last equality is from Condition \ref{c3}.
Thus, Condition C.4 holds under the simple random sampling without replacement.\\
\textbf{Example 2: Poisson sampling}

From the independence of Poisson sampling, we see that
\begin{align*}
\pi_{ij}=&\pi_i\pi_j;\quad\quad\quad i\neq j=1,\dots,N;\\
\pi_{ijk}=&\pi_i\pi_j\pi_k; \quad\quad i\neq j\neq k=1,\dots,N;\\
\pi_{ijkl}=&\pi_i\pi_j\pi_k\pi_l; \quad i\neq j\neq k\neq l=1,\dots,N.
\end{align*}
Hence, $\pi_{ijk}-\pi_{ij}\pi_k=0,$ and $\pi_{ijkl}-4\pi_{ijk}\pi_{l}+6\pi_{ij}\pi_{k}\pi_{l}-3\pi_{i}\pi_{j}\pi_{k}\pi_{l}=0$.
It follows that Poisson sampling satisfies Condition \ref{c4}.

\subsection{A lemma for proving Theorem \ref{them:r} }\label{plem2}
{\lemma\label{lem2} For the classical HT estimator $\hat{\bar{t}}_{HT}$ and the IHT estimator $\hat{\bar{t}}_{\text{IHT}}$, under the conditions \ref{c1}-\ref{c4}, we have
$$E(\hat{\bar{t}}_{HT}-\bar{t})^4=O(n^{-2}), \text{and } E(\hat{\bar{t}}_{\text{IHT}}-\bar{t})^4=O(n^{-2}).$$
}
\begin{proof}
Noting that
 $$\hat{\bar{t}}_{HT}-\bar{t}=\frac{1}{N}{\sum}_U \frac{I_k-\pi_k}{\pi_k}y_k\triangleq\frac{1}{N}{\sum}_U J_ky_k,$$
we have
 \begin{align}
(\hat{\bar{t}}_{HT}-\bar{t})^4&=\frac{1}{N^4}\sum_k\sum_l\sum_i\sum_j (J_ky_k)(J_ly_l)(J_iy_i)(J_jy_j)\nonumber\\
 &=\frac{1}{N^4}\sum_U(J_ky_k)^4
+\frac{4}{N^4}\sum_{k\neq l}(J_ky_k)^3(J_ly_l)
+\frac{3}{N^4}\sum_{k\neq l}(J_ky_k)^2(J_ly_l)^2\nonumber\\
&+\frac{6}{N^4}\sum_{i\neq k\neq l}(J_iy_i)^2(J_ky_k)(J_ly_l)
 +\frac{1}{N^4}\sum_{i\neq j\neq k\neq l}(J_iy_i)(J_jy_j)(J_ky_k)(J_ly_l)\nonumber\\
 &\triangleq \text{\Rma{1}}+\text{\Rma{2}}+\text{\Rma{3}}+\text{\Rma{4}}+\text{\Rma{5}}.\nonumber
 \end{align}
For the first term \Rma{1}, using $\lambda\leq\pi_k\leq 1$ and $|I_k-\pi_k|\leq 1$ for any $k\in U$,  we get
 \begin{align*}
\mid E(I) \mid = E\left(\frac{1}{N^4}\sum_U(J_ky_k)^4\right)
=\frac{1}{N^4}\sum_U\left(\frac{y_k}{\pi_k}\right)^4 E(I_k-\pi_k)^4
\leq \frac{1}{N^4}\sum_U\left(\frac{y_k}{\pi_k}\right)^4=O(n^{-2}).
 \end{align*}
 Similarly, for the terms  \Rma{2} and \Rma{3}, we have
  \begin{align*}
\mid E(J_k^3J_l)\mid=\left|\frac{1}{\pi_k^3\pi_l}E\left[(I _k-\pi_k)^3(I_l-\pi_l)\right]\right|\leq\frac{1}{\pi_k^3\pi_l}E\left[\mid I_k-\pi_k\mid^3\mid I_l-\pi_l\mid\right]\leq\frac{1}{\pi_k^3\pi_l}\leq\frac{1}{\lambda^4},
 \end{align*}
 and
  \begin{align*}
 E(J_k^2J_l^2)=\frac{1}{\pi_k^2\pi_l^2}E\left[(I_k-\pi_k)^2(I_l-\pi_l)^2\right]\leq\frac{1}{\pi_k^2\pi_l^2}\leq\frac{1}{\lambda^4}.
 \end{align*}
Thus, $\mid E(\Rma{2}) \mid =O(n^{-2})$ and $\mid E(\Rma{3}) \mid =O(n^{-2})$.
 \\
Under the conditions \ref{c1} and \ref{c4}, it can be seen that
\begin{align*}
\mid E(J_i^2J_kJ_l)\mid&=\frac{1}{\pi_i^2\pi_k\pi_l}\left|E[(I_i-\pi_i)^2(I_k-\pi_k)(I_l-\pi_l)]\right|\\
&=\frac{1}{\pi_i^2\pi_k\pi_l}\left|E[I_i^2(I_k-\pi_k)(I_l-\pi_l)]-2\pi_iE[I_i(I_k-\pi_k)(I_l-\pi_l)]+\pi_i^2E[(I_k-\pi_k)(I_l-\pi_l)]\right|\\
&=\frac{1}{\pi_i^2\pi_k\pi_l}\left|(1-2\pi_i)E[I_i(I_k-\pi_k)(I_l-\pi_l)]+\pi_i^2(\pi_{kl}-\pi_k\pi_l)\right|\\
&=\frac{1}{\pi_i^2\pi_k\pi_l}\left|(1-2\pi_i)\big[(\pi_{ikl}-\pi_{ik}\pi_l)-\pi_k(\pi_{il}-\pi_i\pi_l)\big]+\pi_i^2(\pi_{kl}-\pi_k\pi_l)\right|\\
&=O(n^{-1}),
\end{align*}
which shows that $\mid E(\Rma{4}) \mid =O(n^{-2})$.\\
Finally, for the term \Rma{5}, we have
\begin{align*}
E\left(\sum_{i\neq j\neq k\neq l}(J_iy_i)(J_jy_j)(J_ky_k)(J_ly_l)\right)&=\sum_{i\neq j\neq k\neq l}\frac{E[(I_i-\pi_i)(I_j-\pi_j)(I_k-\pi_k)(I_l-\pi_l)]}{\pi_i\pi_j\pi_k\pi_l}y_iy_jy_ky_l\\
&=\sum_{i\neq j\neq k\neq l}\frac{\pi_{ijkl}}{\pi_i\pi_j\pi_k\pi_l}y_iy_jy_ky_l-4\sum_{i\neq j\neq k\neq l}\frac{\pi_{ijk}\pi_l}{\pi_i\pi_j\pi_k\pi_l}y_iy_jy_ky_l\\
&+6\sum_{i\neq j\neq k\neq l}\frac{\pi_{ij}\pi_k\pi_l}{\pi_i\pi_j\pi_k\pi_l}y_iy_jy_ky_l-3\sum_{i\neq j\neq k\neq l}\frac{\pi_i\pi_j\pi_k\pi_l}{\pi_i\pi_j\pi_k\pi_l}y_iy_jy_ky_l\\
&=\sum_{i\neq j\neq k\neq l}\frac{\pi_{ijkl}-4\pi_{ijk}\pi_{l}+6\pi_{ij}\pi_{k}\pi_{l}-3\pi_{i}\pi_{j}\pi_{k}\pi_{l}}{\pi_i\pi_j\pi_k\pi_l}y_iy_jy_ky_l.
\end{align*}
Using the conditions \ref{c1} and \ref{c4}, we get
\begin{align*}
\frac{1}{N^4}E\left(\sum_{i\neq j\neq k\neq l}(J_iy_i)(J_jy_j)(J_ky_k)(J_ly_l)\right)&=\frac{1}{N^4}\sum_{i\neq j\neq k\neq l}\frac{\pi_{ijkl}-4\pi_{ijk}\pi_{l}+6\pi_{ij}\pi_{k}\pi_{l}-3\pi_{i}\pi_{j}\pi_{k}\pi_{l}}{\pi_i\pi_j\pi_k\pi_l}y_iy_jy_ky_l\\
&=O(n^{-2}).
\end{align*}
Thus, $E(\hat{\bar{t}}_{HT}-\bar{t})^4=O(n^{-2})$ holds.\\
Similarly, using $\lambda\leq\pi_k^*\leq1$, it is easy to obtain
\begin{align}\label{t4}
E\left(\frac{1}{N}\sum_U \frac{I_k-\pi_k}{\pi_k^*}y_k\right)^4=O(n^{-2}).
\end{align}
In the following, we shall prove $E(\hat{\bar{t}}_{\text{IHT}}-\bar{t})^4=O(n^{-2})$. Noting that
$$\hat{\bar{t}}_{\text{IHT}}-\bar{t}=\frac{1}{N}\sum_U \frac{I_k-\pi_k^*}{\pi_k^*}y_k=\frac{1}{N}\sum_U \frac{I_k-\pi_k}{\pi_k^*}y_k+\frac{1}{N}\sum_U \frac{\pi_k-\pi_k^*}{\pi_k^*}y_k\triangleq A+\Delta,$$
we have
\begin{align}\label{last-equ}
    E(\hat{\bar{t}}_{\text{IHT}}-\bar{t})^4=E(A+\Delta)^4=E(A^4)+4\Delta E(A^3)+6\Delta^2E(A^2)+4\Delta^3E(A)+\Delta^4.
\end{align}
Since $E(A^4)=O(n^{-2})$ from Eqn. (\ref{t4}), we have that $E(A^2)=O(1)$ and $E(A^3)=O(n^{-1})$.
Noting $E(A)=0$ and
$$\Delta=\frac{1}{N}\sum_U \frac{\pi_k-\pi_k^*}{\pi_k^*}y_k=\frac{K}{N}\left(\frac{1}{K}\sum_{U_2} \frac{\pi_k-\pi_k^*}{\pi_k^*}y_k\right)=O(n^{-1}).$$
Therefore, from Eqn.(\ref{last-equ}), we prove that $E(\hat{\bar{t}}_{\text{IHT}}-\bar{t})^4=O(n^{-2})$.
\end{proof}

\end{document}